\input{aipcheck}

\documentclass[
    ,final            
  ]
  {aipproc}

\layoutstyle{8x11double}

\newcommand{\chan}{{\em Chandra}}
\newcommand{\xmm}{{\em XMM-Newton}}

\newcommand{\vlt}{{\em VLT}}
\newcommand{\lbt}{{\em LBT}}
\newcommand{\ntt}{{\em NTT}}
\newcommand{\gem}{{\em Gemini}}
\newcommand{\hst}{{\em HST}}

\newcommand{\spitz}{{\em Spitzer}}

\def \zerofour{RX J0420.0$-$5022}

\begin{document}

\title{Multi-wavelength observations of isolated neutron stars}


\author{Roberto P. Mignani}{
  address={MSSL-UCL, Holmbury St. Mary, Dorking, Surrey, RH5 6NT, UK
}
}

\begin{abstract}
 Forty years elapsed since the optical identification of the first isolated neutron star (INS),  the Crab pulsar.  25 INSs have been now identified in the optical (O), near-ultraviolet (nUV), or near-infrared (nIR), hereafter UVOIR, including rotation-powered pulsars (RPPs), magnetars, and X-ray-dim INSs (XDINSs), while deep investigations have been carried out for compact central objects (CCOs), Rotating RAdio transients (RRATs), and high-magnetic field radio pulsars (HBRPs). In this review I describe the status of UVOIR observations of INSs, their emission properties, and I present recent results. 
 
\end{abstract}

\maketitle


\section{Introduction}

Interestingly enough, although isolated neutron stars (INSs) were discovered in the radio band as sources of strongly beamed radiation (Hewish et al. 1968), hence dubbed {\em pulsars},  one object of this class had been already observed for many years back in the optical band. This is the Baade's star, a quite bright object ($V=16.6$) located at the centre of the Crab Nebula in the Taurus constellation. It was only after the discovery of a bright radio pulsar (NP 0532) at the centre of the Crab Nebula (Comella et al. 1969) that the Baade's star was indeed recognised to be a possible INS and certified as such by the discovery of optical pulsations at the radio period (Cocke et al. 1969). 

Almost a decade passed before the optical identification of another INS: the Vela pulsar (PSR\, B0833$-$45),  one of the faintest objects ever detected on photographic plates ($V=23.6$; Lasker 1976), soon after  identified as an optical pulsar (Wallace et al. 1977). While the search for optical counterparts of radio pulsars continued through the years and scored several new identifications  (see, e.g. Mignani et al. 2004 for a review),  other classes of typically radio-silent  INSs were progressively discovered at X-ray and $\gamma$-ray energies, which also become interesting targets for optical, as well as near-ultraviolet (UV) and near-infrared (IR)  observations.  These are the Soft Gamma-ray Repeaters (SGRs) andf the Anomalous X-ray Pulsars (AXPs), the {\em magnetar} candidates (see Mereghetti 2008; 2009), the X-ray Dim INSs (XDINSs; Haberl 2007; Tr\"umper 2009), and the central compact objects (CCOs) in SNRs (De Luca 2008).  At variance with radio pulsars which are powered by the neutron star rotation (Gold 1968; Pacini 1968), and then are usually referred as rotation-powered pulsars (RPPs),  these INSs radiates through different emission mechanisms like, e.g.  the decay of an hyper-strong magnetic field, for the magnetars, or the cooling of the neutron star surface, for the XDINSs.  At the same time, peculiar radio-loud INSs  were also discovered, like the  high-magnetic field radio pulsars (HBRPs; Camilo et al. 2000) and the Rotating RAdio Transients (RRATs;  Mc Laughlin 2009). In many cases, optical or nIR counterparts have been identified both for the XDINSs and for the magnetars.  

Thus, despite of their intrinsic faintness  which makes them very elusive targets, nUV, optical, and nIR (hereafter UVOIR) observations proved quite successful in detecting INSs outside the radio band, thanks to both the \hst\ and to the new  8m-class telescopes, like the \vlt.  Forty years after the optical identification of the Crab pulsar, UVOIR studies of INSs represent an important tile to complete an understand the multi-wavelength phenomenology of these objects.  In this paper, I will review the status of the UVOIR observations of all different classes of INSs, their emission properties, and the results of recent observations.

\section{RPPs}

Being the first class of INSs detected at UVOIR wavelengths,  the major observational effort has been devoted so far to the search and to the study of RPPs which, indeed, score the largest number of INS identifications.  

After the identification of the Crab and Vela pulsars, the major advances in the UVOIR studies of INSs were achieved thanks to  observations performed with the ESO  \ntt\ in the late 1980s and in the 1990s (see, e.g.  Mignani et al. 2000 for a review).   These yielded to the identification of  the optical counterparts to the first radio-silent  INS, Geminga (Bignami et al. 1993),  to the first extra-galactic INS,  PSR\, B0540$-$69 in the LMC (Caraveo et al. 1992),  and to   PSR\, B0656+14  (Caraveo et al. 1994). The refurbishment of the \hst\ in 1993 represented a major turn-off in UVOIR studies of INSs (see, e.g.  Mignani 2007), yielding to the nUV identification of PSR\, B0950+08, PSR\,  B1929+10 (Pavlov et al. 1996),  PSR\, B1055$-$52 (Mignani et al. 1997), and of PSR\, J0437$-$4715, the first  ms-pulsar  detected at UVOIR  wavelengths (Kargaltzev et al. 2004). Moreover, \hst\ observations allowed to study, for the first time,  the morphology and evolution  of pulsar-wind nebulae (PWNe) around the Crab pulsar (Hester et al. 1995) and PSR\, B0540$-$69 (Caraveo et al. 2000).  In the 2000s, observations with the \vlt\ yielded to the identification of the optical counterparts to PSR\, B1509$-$58 (Wagner\& Seifert 2000) and, more recently, to  PSR\, B1133+16 (Zharikov et al. 2008) and PSR\, J0108$-$1431 (Mignani et al. 2008a).   
A possible counterpart to PSR\, J0108$-$1431 was indeed spotted a few years before by  Mignani et al. (2003), barely detected against the halo of a nearby elliptical galaxy,  but it was not recognised as such until recent  \chan\  observations discovered X-ray emission from the pulsar (Pavlov et al. 2009) and measured its proper motion ($\mu = 200 \pm 65$ mas yr$^{-1}$) through the comparison with the radio position. This allowed Mignani et al. (2008a) to find that the position of the  candidate counterpart ($U=26.4$) nicely fitted  the backward proper motion extrapolation of the pulsar and, thus, to certify its identification {\it a posteriori}

For many objects, the faintness of the optical counterpart ($V \ge 25$) initially prevented the use of the straightforward optical timing identification technique, successfully used for the brighter Crab and Vela pulsars and for PSR\, B0540$-$69 (Shearer et al. 1994). In some cases,   the proper   motion measurement of the  candidate  counterpart, successfully tested for Geminga  (Bignami et al.  1993), was used instead,  confirming the identification of PSR\, B0656+14  (Mignani et al. 2000) and of PSR\, B1929+10 (Mignani et al. 2002), both achieved through high-resolution \hst\  imaging.  For both Geminga and PSR \, B0656+14, the identification was later strengthened  by the detection of nUV pulsations with the \hst\ (Kargaltsev \& Pavlov 2007). For the others, the identification evidence mainly relies on the positional coincidence with the INS coordinates (PSR\, B1133+16 and PSR\, J0108$-$1431), on the peculiar colours or spectrum of the candidate counterpart (PSR\, B0950+08), on the possible evidence of optical polarisation (PSR\, B1509$-$58). 

Recently, the optical identification of PSR\, B1055$-$52 was confirmed through new \hst\ observations (Mignani et al. 2009a). The candidate counterpart was clearly detected in the  nUV with the {\em ACS/SBC} (Fig. 1) and in the optical with the {\em WFPC2}.  Together with the original $U$-band {\em HST/FOC} photometry of Mignani et al. (1997), the new \hst\ flux measurements allowed to measure the object peculiar colours which virtually certify it as the actual counterpart of PSR\, B1055$-$52. Moreover, the comparison between the relative position of the counterpart, as measured in the 1996 {\em FOC} observations and in the new {\em ACS} ones,  allowed to measure a proper motion of $\sim$ 55 mas yr$^{-1}$,  the first ever measured for this pulsar at any wavelength. This corresponds to a tangential velocity $v_{T}  \sim 180  d_{700}$, where $d_{700}$ is the source distance in units of 700 pc.
Thus, out of the $\sim 1800$ RPPs listed in the updated ATNF radio catalogue, twelve have been identified at UVOIR wavelengths,  albeit with different levels of confidence. 

\begin{figure}
 \includegraphics[height=7cm]{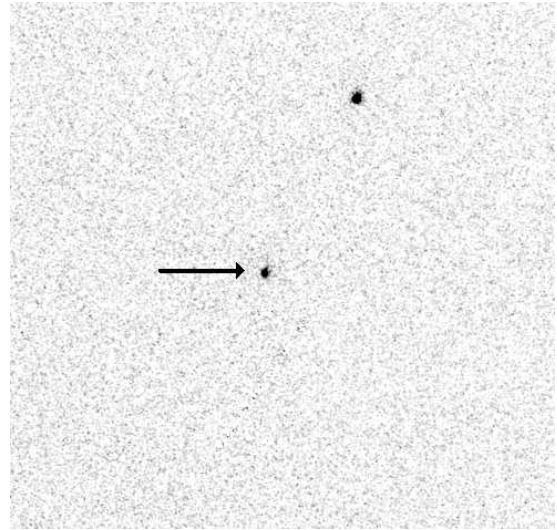}
  \caption{\hst\  {\em ACS/SBC} image of the  PSR\, B1055$-$52 field  ($12" \times 12"$). North to the top, East to the left.   The pulsar counterpart is marked by the arrow. The star $\sim$ 4" north of the pulsar is star A of Mignani et al. (1997). }
\end{figure}

The study of the UVOIR emission properties of RPPs is complicated by the paucity of spectral information. Only four RPPs have a complete UVOIR spectral coverage, with the nUV and nIR spectral bands crucial to disentangle the contributions of thermal and non-thermal emission processes. Moreover, only for five RPPs low-resolution spectroscopy is available, while for the others the knowledge of the spectral energy distribution (SED) still relies on, sometimes sparse, multi-band photometry, with all the implied caveats (see, e.g. discussion in Mignani et al. 2007a).  Only five RPPs pulsate in the optical/UV, where they all feature double-peaked light curves, with the exception of PSR \,B0540$-$69, and robust optical polarisation measurements exist only for the Crab pulsar (see, e.g. Slowikowska et al. 2009 and references therein), reducing the impact of these important diagnostic tools. 
The SED of the RPPs grows in  complexity with the spin-down age (see, e.g. Fig. 2 of Mignani et al. 2007a). It evolves from a single power-law  (PL) spectrum ($F_{\nu} \propto \nu^{-\alpha}$; $\alpha \sim 0$) for young RPPs (age$< 10^{4}$ years)  to  a  composite  one  featuring both  a PL  ($\alpha \ge 0$) and blackbody (BB) component ($T\sim 2-8 \times 10^{5}$ K) for the middle-aged ones (age $10^{5}-10^{6}$ years), the former dominating in the IR, the latter in the UV.  A similar composite SED was measured also for the middle-aged PSR\, B1055$-$52 (Mignani et al. 2009a). Surprisingly, older RPPs  (age$>10^{6}$ years) seem to feature single PL spectra.  For the very old ($10^{9}$ years) PSR\, J0437$-$4715,  the nUV spectrum is consistent with  a pure BB (Kargaltsev et al. 2004). The PL component is ascribed to non-thermal radiation produced by relativistic particles accelerated in the neutron star magnetosphere (e.g., Pacini \& Salvati 1983).   On the other hand, the BB component is ascribed to thermal emission produced from the cooling neutron star surface.
In general, there  is a  correlation between the optical luminosity and the neutron star rotational energy loss (Zharikov et al. 2006), suggesting that the magnetospheric emission is powered by the star rotation and contributes most to the optical luminosity. Emission efficiencies $L_{opt}/\dot{E} \approx 10^{-6}-10^{-7}$ are  then derived. It is interesting to note that the BB/PL  components which fit the RPP optical spectra are not consistent with the extrapolation of the analogue BB/PL components which fit the X-ray spectra (Mignani et al. 2004). The presence  of two PL components clearly indicates a break in the magnetospheric spectrum.  No evident correlation is found between the optical PL spectral index and the neutron star age (Mignani et al. 2007a), like none is found in the X-rays (Becker et al. 2009).  On the other hand, the presence of distinct optical and X-ray BB components suggests that the temperature distribution on the neutron star surface is not homogeneous.  

 Interestingly, Kargaltsev et al. (2004) noted that the BB temperature ($T\sim 2 \times10^{5}$ K)  derived for PSR\, J0437$-$4715  is larger than expected from standard cooling models. This suggests that either the neutron star surface temperature was raised through the effect of re-heating processes occurring in its  interior  (e.g., Page 2009; Tsuruta 2009) or it was raised by accretion from its companion star during a past pulsar spin-up phase.  Like for  the old ms-pulsar PSR\, J0437$-$4715, re-heating might have occurred  also for the 200 million year old PSR\, J0108$-$1431.  The accurate measurement of its distance (240 pc), obtained through radio parallax measurements (Deller et al. 2009), implies that the optical fluxes could be compatible with thermal emission from the bulk of the neutron star surface at a BB temperature of $\approx 3 \times 10^{5}$ K.  Again, this temperature is larger than the values expected from standard cooling models.

\section{XDINSS}

Soon after their discovery in the 1990s, XDINSs were the next class of INSs  to attract interest from the neutron star optical astronomy community. Indeed,  without the evidence coming from the detection of X-ray pulsations, detected only later with \xmm\ (e.g., Haberl et al. 2007), the measurement of an extreme X-ray-to-optical flux ratios $F_{X}/F_{opt}$ was, at that time, the only way to certify these sources as INSs. Although no XDINS optical counterpart was initially identified  from \ntt\ observations, despite of a considerable observational effort, they were nonetheless crucial to pave the way to deeper observations. 
The first XDINS to be detected in the optical was RX\, J1856.5$-$3754 which was indeed observed by the \hst\  (Walter \& Matthews 1997). Riding the wave, \hst\  also detected likely optical counterparts  for  RX\, J1308.6+2127 and  RX\, J1605.3+3249 (Kaplan et al.  2002, 2003).  From the ground, optical detections were obtained for RX\, J0720.4$-$3125 with the \ntt\ (Motch  \& Haberl  1998) and with the {\em Keck} (Kulkarni \& van Kerkwijk 1998) and for RBS\,1774  with the \vlt\  (Zane  et al.  2008) and the \lbt\ (Schwope et al. 2009). However, the optical identifications were certified by proper motion measurements only for three of them: RX\, J1856.5$-$3754 (Walter et al. 2001), RX\, J0720.4$-$3125 (Motch et al.  2003),  and RX\, J1605.3+3249 (Motch et al. 2005; Zane et al. 2006),  while for  both RX\, J1308.6+2127 and RBS\, 1774 the optical identifications are still based on the positional  coincidence with the \chan\ X-ray coordinates  and on the $F_{X}/F_{opt}$ ratio.   For the two XDINSs for which optical parallaxes were measured with the \hst, RX\, J1856.5$-$3754  and   RX\, J0720.4$-$3125 (van Kerkwijk \& Kaplan  2007), the proper motion measurement also allowed to derive their space velocities and, thus, to rule out accretion from the interstellar medium (ISM).  

Recently, a possible candidate counterpart was identified for RX\, J0420.0$-$5022.  Originally, a candidate counterpart  was tentatively proposed  by Haberl et al.  (2004), based on \vlt\ observations. However,  a re-analyses of their  \vlt\ data (Mignani et al. 2009b) showed that their candidate was detected only at the  $\sim 2 \sigma$  level and it was most likely a spurious detection produced by the halo of a very bright ($B=10$)  star located 40" away.  At the same time, deeper \vlt\ observations  (Mignani et al. 2009b) allowed to find evidence for the detection ($\sim 4  \sigma$) of a fainter object ($B= 27.5\pm 0.3$) coincident with the \chan\ position of \zerofour\ (Fig. 2).

\begin{figure}
\includegraphics[height=7cm,clip]{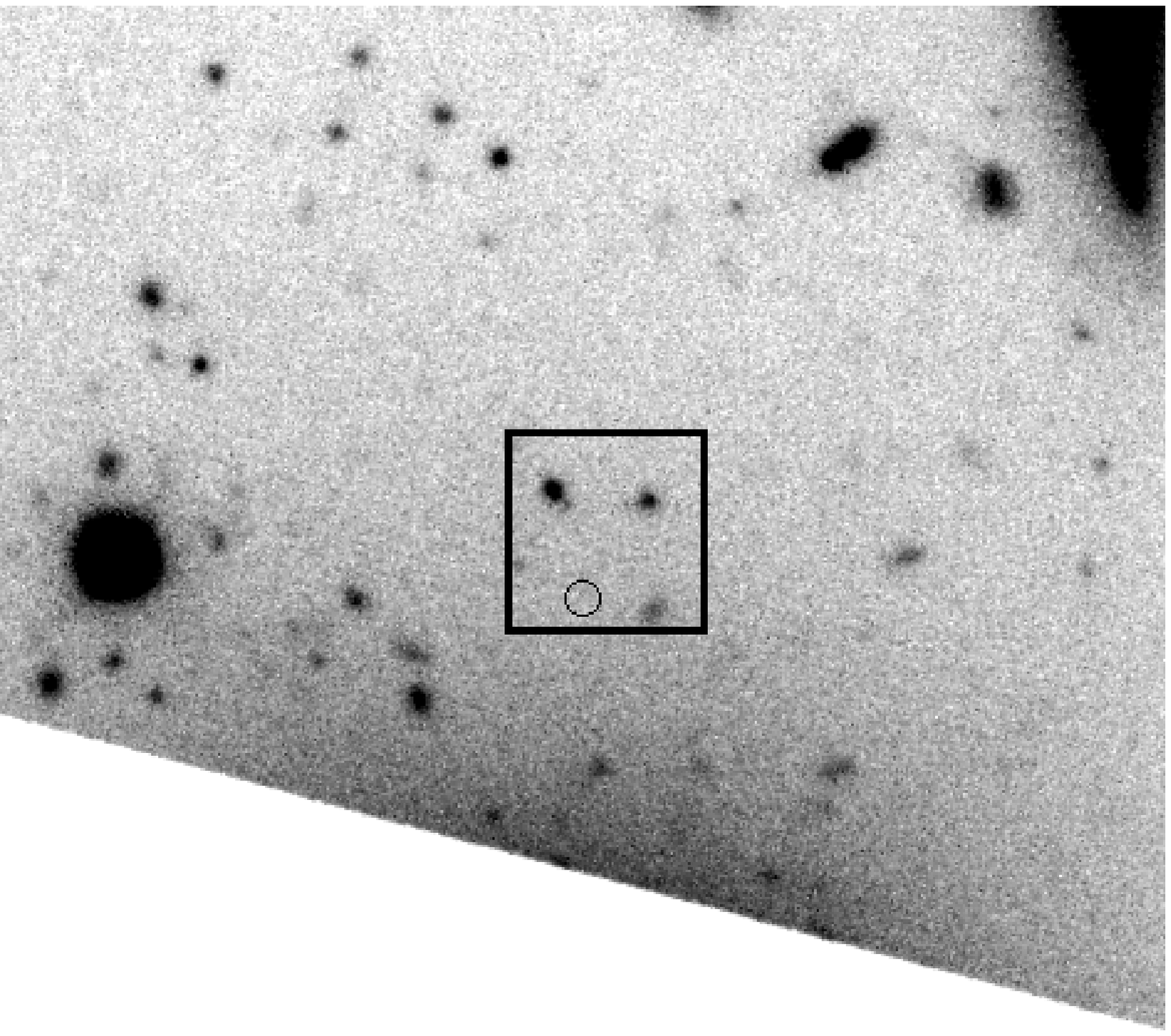} 
\includegraphics[height=7cm,angle=0,clip]{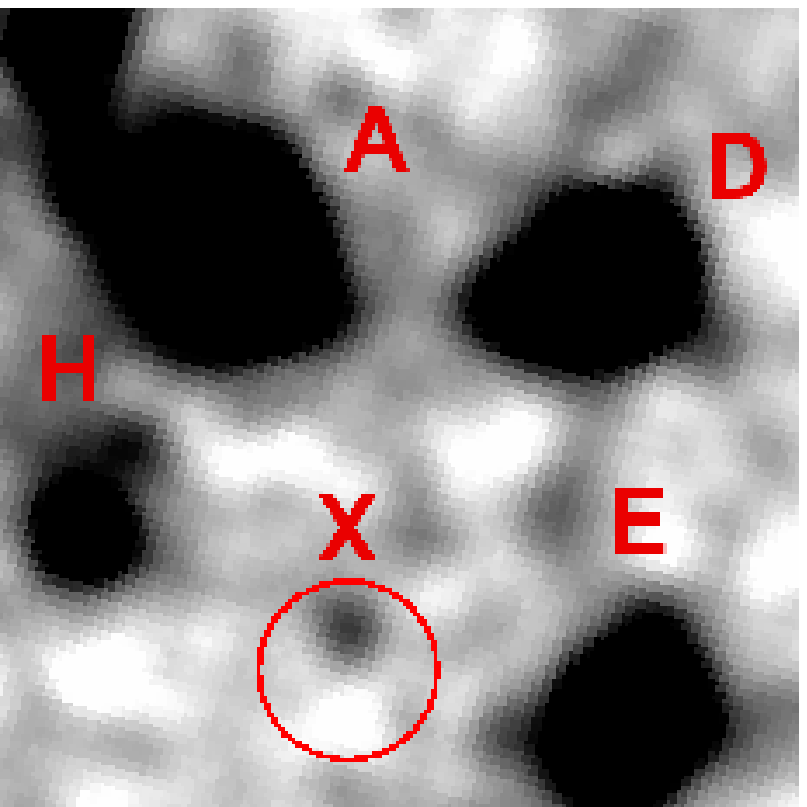}
  \caption{Left: \vlt/{\em FORS2}  B-band image  ($1'\times 1'$)  of the \zerofour\ field. North  to the top, east to the  left. The circle (1.11" radius; 90 \% confidence level)  indicates the   \zerofour\ position  (Mignani et al. 2009b). A bright star south of \zerofour\ is masked.  Right: $10"\times 10"$ zoom of the same region smoothed with a Gaussian filter over cells of  $5\times5$ pixels. Object $X$  ($B= 27.5\pm 0.3$)  is the putative \zerofour\ counterpart. Field objects are labelled as in Haberl et al. (2004). }
\end{figure}

XDINSs are only detected at optical wavelengths. Although they have been repeatedly observed in the nIR with the \vlt, none of them was detected so far (Mignani et al. 2007b, 2008b; Lo  Curto et al.  2007; Posselt et al. 2009).  No observation of XDINSs in the nUV was reported yet. Like for the RPPs, the knowledge of the SED mainly relies on multi-band photometry, with low-resolution spectroscopy only performed for RX\, J1856.5$-$3754 (van  Kerkwijk \& Kulkarni 2001).  No search for optical pulsations or polarisation measurements have been carried out so far.

The XDINS optical fluxes usually exceed by a factor  of $\sim 5$ (or more) the  extrapolation of the X-ray blackbody (see, e.g. Kaplan 2008).  For the XDINSs with an inferred rotational energy loss, i.e. RX\, J0720.4$-$3125,  RX\, J1308.6+2127, RX\,  J1856.5$-$3754, RBS\, 1774 (Kaplan \& van Kerkwijk 2005a,b; van Kerkwijk \& Kaplan 2008; Kaplan \& van Kerkwijk 2009) this turns out to be too low ($\approx 10^{30}$ erg s$^{-1}$)  to sustain magnetospheric emission powered by the star rotation, unless the emission efficiency is a few orders of magnitude higher than in RPPs.  In at  least in the best-studied  cases of RX\, J1856.5$-$3754 (van  Kerkwijk \& Kulkarni 2001) and  RX\, J0720.4$-$3125 (Motch et al. 2003), the optical spectra is found to closely follow a  Rayleigh-Jeans, which suggests that the optical emission is predominantly thermal ($T\sim 2-4 \times10^{5}$ K).  The XDINS optical emission has been thus interpreted either in terms of a non homogeneous surface temperature distribution, with the larger and cooler part emitting the optical  (e.g., Pons et al. 2002),  or of reprocessing of the surface radiation by a thin H atmosphere around a bare neutron star  (Zane et al. 2004; Ho 2007). For RX\, J1605.3+3249,  multi-band photometry measurements do exist (Motch et al. 2005) but they are likely affected by instrument cross-calibration problems which hamper the SED characterisation (Zane et al. 2006). For RXJ\, 1308.6+2127 (Kaplan et al. 2002), RBS\, 1774 (Zane et al. 2008; Schwope et al. 2009), and  \zerofour\  (Mignani et al. 2009b) only one band measurement is available

Interestingly,  RBS\, 1774 is characterised by an anomalously large optical excess of $\approx 35$ (Zane et al. 2008). This is unlikely produced from the cooling neutron star surface unless, to explain the low observed pulsed fraction of the X-ray light curve,  one invokes a peculiar alignment between the magnetic field,  the rotations axis, and the line of sight (Zane et al. 2008).   The RBS\, 1774 emission may thus be of magnetospheric origin,  perhaps related to its large magnetic field ($\sim 10^{14}$ G), inferred from the observations of an absorption feature in the X-ray spectrum (Zane et al. 2005).  However, the recent measurement of the pulsar spin-down (Kaplan \& van Kerkwijk 2009) implies a magnetic field of $\sim 2 \times 10^{13}$ G, thus possibly arguing against this interpretation. A recent re-analysis of the \xmm\  spectrum of RBS\, 1774 using a different spectral model (Schwope et al. 2009), while confirming the large optical excess,  does not rule out the possibility that the optical emission is indeed thermal.   For \zerofour, the flux of the putative counterpart corresponds to an optical excess of $\sim 7$ with respect to the extrapolation of the \xmm\ spectrum (Mignani et al. 2009b).
From  the upper limit on the \zerofour\ rotational energy loss, $\dot {E} < 8.8 \times 10^{33}$ erg s$^{-1}$ derived from X-ray timing (Haberl et al. 2007), the optical luminosity of the putative counterpart, $L_{B} \sim 1.2 \times 10^{27}$  erg s$^{-1} ~ d_{350}^2$ (where  $d_{350}$ is the distance in units of 350 pc; Posselt et al. 2007), implies an  emission efficiency $> 1.3 \times 10^{-7}$ for rotation-powered emission. This value could still be compatible with the range of emission efficiencies derived for $10^{6}-10^{7}$ years old RPPs (Zharikov et al. 2006).  In case of thermal emission from the neutron star surface, the optical luminosity of the putative counterpart would be compatible with a blackbody with temperature $\leq 25$ eV and an  an implausibly large emitting radius of $\geq$ 23 km.  Thus, would the putative counterpart be confirmed, its optical luminosity, for a neutron star distance of $\sim 350$ pc,  might rather point towards a non-thermal origin for the optical emission.

\section{sgrs and axps}

\begin{figure}
 \includegraphics[height=7cm]{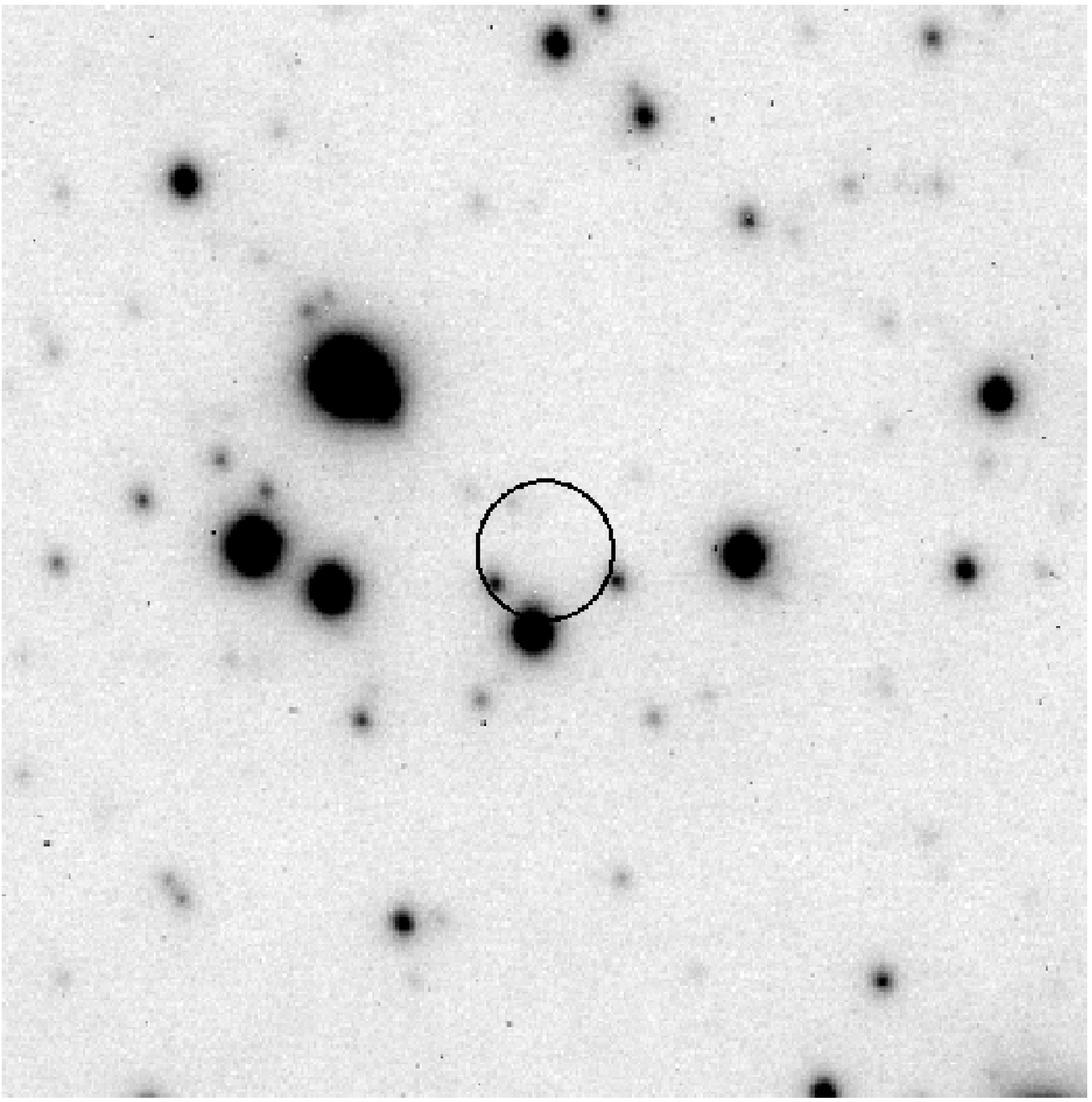}
 \includegraphics[height=7cm]{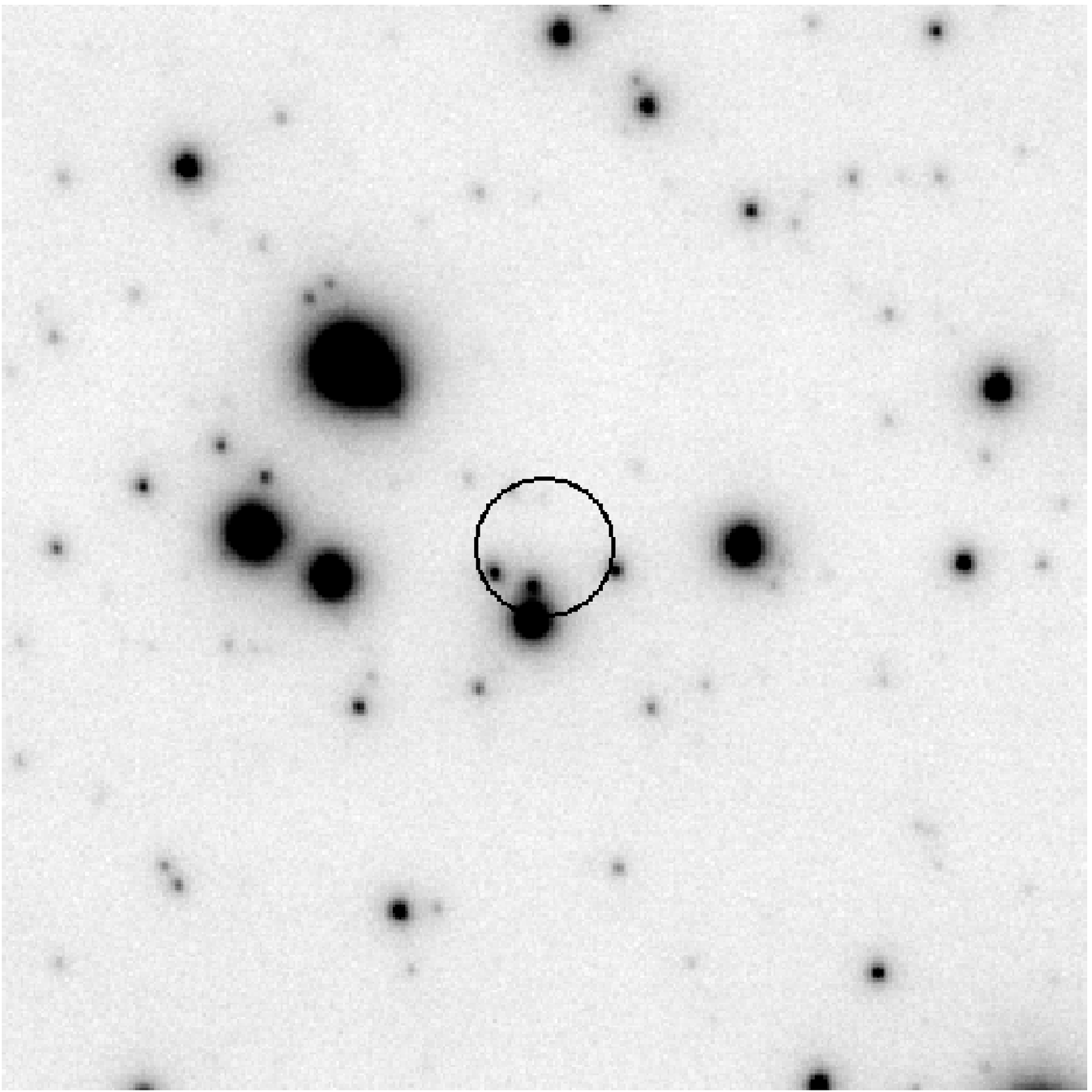}
  \caption{\vlt/{\em NACO} $K_{s}$ band images ($10"\times10"$) of the AXP 1E\,1540.0$-$5408 field obtained in July 2007 (left; Mignani et al. 2009c) and in January 2009  (right; Israel et al. 2009a). North to top, East to the left. The variable object in the computed $3\sigma$ radio error circle (0.63") is the nIR counterpart.}
\end{figure}

Optical/IR observations of AXPs and SGRs, the magnetar candidates, were originally carried out  (e.g.,  Davies  \& Coe 1991; van Paradijs et al. 1996) to investigate the nature of these sources and to test models based on accretion from a companion star or a debris disc.  Unfortunately,  optical/nUV observations of magnetars are hampered by the high interstellar extinction (up to $A_{V} \approx 30$) in the direction of the galactic plane,  where most of them were discovered. This makes the nIR the only spectral band suitable for observations, although they are complicated by the field crowding. Indeed, nIR identifications of magnetars were boosted when deep adaptive optics (AO) observations became possible with 8m-class telescopes like the \vlt,  \gem, as well as with the  {\em CFHT}, allowing to perform high-resolution imaging and to resolve potential candidate counterparts within the \chan\ X-ray source error circles.  However, unless a more accurate source position is available from radio observations like, e.g. in the case of the two transient radio AXPs XTE\, J1810$-$197 and 1E\,1540.0$-$5408 (Camilo et al. 2007a,b) and of SGR\, 1806$-$20 (Cameron et al. 2005), the positional coincidence alone is not sufficient to pinpoint the actual counterpart. Given the source distance and the longer required time span, proper motion measurements are not a very efficient  identification tool like, e.g. in the case of RPPs and XDINSs, while the search for nIR pulsations at the X-ray period requires suitable instruments which are usually not available at most telescopes. Indeed, optical pulsations have been detected from two AXPs (see below) using guest instrument facilities. Moreover,  the still significant interstellar extinction, together with the uncertainty on the source distance and the difficulty of finding a well-defined spectral template for magnetars, makes a colour-based identification uncertain.  Thus, in most cases the best recipe for nIR identifications of magnetars is based  on the measurement of correlated IR/X-ray variability.  The application of this recipe, of course, benefited from the possibility of performing prompt Target of Opportunity (ToO) observations with ground-based telescopes in response of triggers from high-energy satellites. 

Out of the five currently  known SGRs,  SGR\, 1806$-$20 has been the first to be firmly identified in the nIR (Israel et al.  2005; Kosugi et  al.  2005). Recently, a second nIR identification has been obtained for the newly discovered  SGR\, 0501+4516 thanks to prompt follow-up observations with the {\em UKIRT}, the {\em WHT}, and the {\em Gemini} (Rol et al. 2009) which pinpointed a $K_{s}\sim 19.1$ counterpart at the boresight-corrected 0.1" \chan\ position (Woods et al. 2008). The same observations also showed a possible evidence for nIR variability, although not evidently correlated with that observed in the X-ray band. The counterpart was also detected in the optical (see Rol et al. 2009 et references therein). At the same time, four out of the eight known AXPs  have been identified in the nIR:  4U\, 0142+61 (Hulleman et  al.  2004), 1E\,  1048.1$-$5937 (Wang  \& Chakrabarty  2002; Israel et  al.  2002), XTE\, J1810$-$197  (Israel et  al.  2004; Rea  et al. 2004),  and 1E\, 2259+586 (Hulleman  et al.  2001).  In addition, both 4U\, 0142+61 (Hulleman et  al.  2000) and 1E\,  1048.1$-$5937 (Durant \& van Kerkwijk 2005) have  been identified in the optical.  4U\, 0142+61 and  1E\, 2259+586 are the only AXPs/SGRs which has been detected  in the mid-IR  by \spitz\ (Wang et  al.  2006; Kaplan et al. 2009), while upper limits have been obtained for XTE\, J1810$-$197, 1RXS\, J170849.0$-$400910 (Wang et al. 2007a) and 1E\,1048.1$-$5937 (Wang et al. 2007a; 2008).    nIR  counterparts were also proposed for  SGR\, 1900+14 and for the AXP 1E\, 1841$-$045  (Testa et  al.  2008), based  on possible  long-term nIR variability.  Optical flares of a possible candidate SGR, SWIFT\, J195509.6+261406,  were detected by Stefanescu et al. (2008) using the {\em OPTIMA} instrument at the 1.3-m telescope of the Skinakas Observatory (Crete), with the possible evidence of periodicity (6-8s) in the strongest flares.

Recently, an nIR counterpart was identified for the radio-transient AXP 1E\,1540.0$-$5408 through \vlt\ observations performed in January 2009, right after the source underwent an X-ray outburst detected by the {\em Swift/BAT} (January 22;  Gronwall et al. 2009).  A candidate counterpart (Fig. 3) was identified coincident with the radio error circle of 1E\,1540.0$-$5408 (Camilo et al. 2007b) with a magnitude $K_{s}=18.5$ (Israel et al. 2009a). The candidate counterpart was not detected in {\em Magellan} observations ($K_{s}>20.1$) obtained just four days before the X-ray outburst (Wang et al. 2009). The candidate counterpart was also not detected in much  deeper \vlt\ images of the same field acquired in July 2007 (Mignani et al. 2009c) down to a limiting magnitude of $K_{s} \sim 21.7$.  This implies a long-term variability of about 3 magnitudes ($K_{s}$ band), which  virtually certifies its identification with 1E\,1540.0$-$5408.  

 It is interesting to note how all magnetar identifications were achieved with ground-based telescopes, mostly with the \vlt, while the \hst\  did not play a significant role, partially because its nIR camera ({\em NICMOS}) was unavailable between 1999 and 2002.

Historically, optical/nIR observations of SGRs and AXPs  were fundamental to provide crucial supporting evidence for the INS scenario and for the magnetar model.  In the IR, magnetars are characterised by a flux excess with respect to the extrapolation of the X-ray spectrum (see, e.g. Israel et al. 2005). Although this has been often referred in the literature to be a distinctive character of magnetars, many RPPs indeed feature a similar excess (see, e.g. Mignani et al. 2007a) and it can be interpreted in terms of either a spectral break or of the onset of a new spectral component.

The  magnetar optical/nIR SED is not fully characterised  yet. Indeed, in most cases the accessible  spectral coverage is limited to the near-IR, while mid-IR observations are hampered by their lower spatial resolutions. In the optical,  only the AXPs 4U\, 0142+61 and 1E\,  1048.1$-$5937, and now SGR\, 0501+4516, were detected, but only the former at wavelengths bluer than the $I$ band. In all cases, the source of spectral information still relies on multi-band photometry. This comes with an important caveat since magnetars are known to be variable in the optical/nIR and thus the available multi-band photometry observations, which are  often taken  at different epochs,  are not directly comparable. 
Due to their faintness, no optical/nIR spectroscopy observations have been performed for any magnetar, so far. The only exception is 1E\,1540.0$-$5408 for which both spectroscopy and polarimetry observations were recently performed to follow-up the identification of the nIR counterpart  (Israel et al. 2009a).  However, at the time of writing of this manuscript, the data analysis is still under way. Like for the RPPs, useful information come from timing and polarimetry observations.  So far, pulsations at the X-ray period were detected in the optical only for the AXPs 4U\, 0142+61 (Kern \& Martin 2002; Dhillon  et al.  2005) and 1E\, 1048.1$-$5937 (Dhillon  et al.  2009). No pulsation was  detected in the nIR for 4U\, 0142+61 down to a pulsed fraction of 17\% (Morii et al. 2009).  Phase-averaged polarimetry observations have been performed in the $K_s$ band with the \vlt\ only for 1E\, 1048.1$-$5937 and XTE\, J1810$-$197 (Israel et al. 2009b, in preparation).

Thus, due to the lack of a well-characterised SED, the origin of the optical/nIR emission of the magnetars is very much debated. While for both RPPs and XDINSs the UVOIR emission comes from the neutron star, for the magnetars it is still not clear yet whether the optical/nIR emission also comes from the neutron star.  Interestingly, Mignani et al. (2007c) showed that, if powered by the neutron star rotation, the magnetar nIR efficiency $L_{IR}/\dot{E}$ would  be at least two orders of magnitudes larger than that of RPPs.  This larger nIR output might be possibly related to the larger magnetar magnetic fields (see, e.g. Fig. 3 of Mignani et al. 2007c) rather than to the neutron star rotation, or to reprocessing of the X-ray radiation in a fallback disc around the magnetar. For the two magnetars with the broadest optical/nIR spectral coverage, 4U\, 0142+61 and 1E\,  1048.1$-$5937, the SED can be fitted by a PL (e.g., Wang et al. 2006; Durant \& van Kerkwijk 2005), which would favour a magnetospheric origin of the optical/nIR emission.  On the other hand, the detection of the AXP  4U\, 0142+61 in the mid-IR (Wang  et al.   2006), where it features a clear spectral bump with respect to the PL which fits the optical/nIR fluxes, has been considered a possible evidence for the presence of a disc. A similar evidence was recently found also from mid-R observations of 1E\, 2259+585 (Kaplan et al. 2009). No other AXP/SGR has been detected in the mid-IR so far (Wang et al. 2007a). For  1E\,1048.1$-$5937 (Wang et al. 2008), the deep \spitz\ upper limits tend to rule out the presence of a disc, at least assuming the same size and geometry of the disc claimed around  4U\, 0142+61. 
 
The comparison between the pulsed fractions and the relative phases of the optical and X-ray light curves  provides an important diagnostic to discriminate between magnetospheric and disc emission. In the case of 4U\, 0142+61,  Dhillon et al. (2005) found that the optical pulsed fraction ($29\% \pm 8\%$) was  $\sim$5 times larger than the X-ray one and found no evidence for a significant optical-to-X-ray pulse lag, which would argue against the optical emission being due to X-ray reprocessing in a disc. However,  the optical and X-ray observations were not simultaneous and the variation of the X-ray pulsed fraction along the source state might have hampered the comparison.  On the other hand, in the case of 1E\, 1048.1$-$5937  Dhillon  et al.  (2009) found,  through simultaneous optical and X-ray observations, that the optical pulsed fraction  ($21\% \pm 7\%$)  was actually a factor $\sim 0.7$ lower than the X-ray one but they also found a possibly significant evidence of an X-ray-to-optical pulse lag (0.06$\pm$0.02). 
All in all,  timing observations tend to favour a magnetospheric origin of the pulsed optical emission.  However, a disc origin of the magnetar optical/nIR emission is  still  considered a possibility (see,  e.g.  Ertan et al.   2007),  at least to explain the unpulsed emission component.  

A magnetospheric origin of the nIR emission is supported by the erratic  nIR-to-X-ray  variability  observed in  XTE\,  J1810$-$197  (Testa et al. 2008), which is hardly compatible with  X-ray reprocessing in a disc. Possible evidence might also come from the nIR-to-X-ray decay rate of  1E\, 1048.1$-$5937 in the 2.5 month following the March 2007 outburst. \vlt\ data (Israel et al. 2009c, in preparation) show evidence for a larger decay in the nIR with respect to the X-rays, which also argues against disc reprocessing. The upper limit of $\sim$ 25\% on the phase-averaged nIR polarisation of 1E\, 1048.1$-$5937 (Israel et al. 2009b), with respect to the $\sim$10\% measured in the optical for RPPs (see, e.g. Slowikowska et al. 2009 and references therein), does not provide compelling evidence for a non-thermal origin of the magnetar nIR emission, though. 
 
 \begin{table}
\begin{tabular}{lllllllll} 
\hline
{Name} & {Age}  & {mag} & {d(kpc)} & {$A_V$}  & {Phot.} & {Spec.} & {Pol.}  & {Tim.} \\ \hline
Crab     	                 & 3.10 &16.6 		    & 1.73  & 1.6    &  nUV, O, nIR & Y & Y  & Y \\
B1509$-$58              & 3.19 & 25.7$^R$  & 4.18  & 5.2    &  O         &     & Y  &  \\
B0540-69				& 3.22 & 22.0 	    & 49.4  & 0.6    &  O         &  Y & Y  & Y        \\
Vela					& 4.05 & 23.6 	    & 0.23  & 0.2    &  nUV, O, nIR & Y & Y  &  Y \\
Geminga				& 5.53 & 25.5 	    & 0.07  & 0.07  &  nUV, O, nIR & Y &     & Y \\
B0656+14				& 5.05 & 25.0 	    & 0.29  & 0.09  &  nUV, O, nIR & Y & Y  & Y\\
B1055$-$52			& 5.73 & 24.9$^U$  & 0.72  & 0.22  &  nUV, O     & 	  & 	    &        \\
B1929+10				& 6.49 & 25.6$^U$  & 0.33  & 0.15  &  nUV     &	  &     &\\
B1133+16                                    & 6.69   & 28 &          0.35      & 0.12   & O   & & & \\
B0950+08				& 7.24 & 27.1 	    & 0.26  & 0.03  &  nUV, O     &    &     &\\
J0108$-$1431                             & 8.3    & 26.4$^U$  & 0.2    & 0.03    &  O          &           &     & \\
J0437$-$4715		& 9.20 &             & 0.14  & 0.11  &             & Y &     &	\\ \hline
RX\, J1308.6+2127	& 6.11 & 28.6$^R$  & $<$1 &    0.14      & O           &    &     & \\ 
RX\, J0720.4$-$3125	& 6.27 & 26.7  	    & 0.30  & -0.3       & nUV, O      &    &     &\\
RX\, J1856.5$-$3754	& 6.60 & 25.7  	    & 0.14  & 0.12  & nUV, O      & Y &	    & 		  \\   
RX\, J1605.3+3249	&		& 26.8$^R$  & $<$1 & 0.06  & O           &    & 	    &  \\
RBS\, 1774                   &		& 27.2$^B$  & 0.34  &  0.18       & O           &    &      &\\ 
RX\, J0420.0$-$5022 &             & 27.5 $^B$  & 0.35 & 0.07         &  O          &     &       &  \\ \hline
1E\, 1547.0-5408		& 3.14 & 18.5          & 9       & 17     & nIR          &    &      &\\
SGR\, 1806$-$20		& 3.14 & 20.1          &15.1   & 29     & nIR         &     &     &  \\ 
1E\, 1048.1$-$5937  & 3.63 & 21.3          &3.0     & 6.10  & O, nIR          &    &  Y  &  Y \\
XTE\, J1810$-$197	& 3.75 & 20.8          &4.0     & 5.1    &  nIR         &    &   & \\
SGR\, 0501+451		& 4.10 & 19.1          &    5     & 5       & O, nIR         &     &     &  \\
4U\, 0142+61            & 4.84 & 20.1          &1.73   & 1.62  & O, nIR,mIR      &    &      & Y \\
1E\, 2259+586		& 5.34 & 21.7          &3.0     & 5.7    & nIR,mIR          &    &      & \\ \hline 
\end{tabular}

\caption{UVOIR identifications of  INSs:  RPPs (first group), XDINSs (second), and magnetars (third). Each group is sorted according to the spin down age. Column 1 gives the INS name, while columns 2 to 5 give the spin down age (in logarithmic units),  magnitude, distance, and interstellar extinction $A_V$.  For RPPs and XDINSs, magnitudes refer to the V band, unless otherwise indicated by the superscript, while for magnetars they refer to the K$_s$ band. Column 6 to 8 give  the broadband photometry coverage in the near-ultraviolet (nUV), optical (O), and near/mid-infrared (nIR,mIR) spectral bands, and a flag (Y) for spectroscopy, polarisation, and timing measurements.   }
\end{table}

\section{CCOs}

For the CCOs,  deep optical/nIR observations have  been performed  for nearly all sources using the \hst, the \vlt, and the \gem. For  the CCO  in PKS  1209$-$51, the originally proposed nIR identification with a low-mass M-star (Pavlov et al.   2004) was ruled out by an improved astrometry analysis  (Mignani et al.   2007b; Wang et al.  2007b)  and by the upper limit on the proper motion of the proposed candidate counterpart  (De Luca et al.  2009).  No viable candidate counterpart was found for the CCOs  in Cas A  (Fesen  et al.  2006),  Kesteven 79 (Gotthelf et al. 2005), G347.3$-$0.5 (Mignani et al. 2008c),  Puppis A (Wang et al.  2007b; Mignani et al. 2009d). For all of them, the derived optical/nIR upper limits enable to rule out the presence of  a binary companion other than  a very low  mass star (M5 type or later), although they do not exclude the presence of a debris disc as well, like in the case of  the magnetars.   Only  for the  CCO  in  Vela Jr.   a possible  nIR counterpart ($H\sim 21.6$;$K_{s}\sim 21.4$) was identified by the \vlt\ (Mignani  et al. 2007d), whose nature is yet undetermined and could be compatible with either being the neutron star itself, a low-mass M-star, or  a fallback disc. Interestingly, the Vela Jr. CCO features an optical nebulosity detected in the $R$ band with the \vlt\  and also detected in the $H_{\alpha}$ line, which has been interpreted as a possible bow-shock, produced by the neutron star motion in the ISM or  a photo-ionisation nebula (Mignani et al. 2007b,d). However, new \hst\ observations (Mignani et al. 2009e) apparently contradicts both scenarios.
For the CCO in RCW 103, often suspected to be in a binary system because of its transient X-ray emission and 6 its hours periodicity,  a candidate counterpart was proposed (Pavlov et al. 2004) with an M-star identified close to the \chan\ position. However, a systematic re-analysis of the astrometry of all the available \chan\ observations of the CCO (De Luca et al. 2008) ruled out the association with the proposed counterpart. A search for correlated X-ray and nIR variability, both along the source 6 hours period and along a time base line of  years, from all possible counterparts detected around the \chan\ position was carried out by De Luca et al.  (2008) using all the available \hst\ and \vlt\ data sets.  However, no significant evidence of variability at any time scale was found.

\section{RRATs and HBRPS}

For the RRATs, a search for bursting optical emission from J1819-1458  were performed by Dhillon et al. (2006) but with negative results. More recently, nIR observations were performed with the \vlt\ for all RRATs which are detected in X-rays, i.e. J1317$-$5759, J1819$-$1458, and J1913+1333. However, no  counterpart was singled out based on colours and/or variability  (Rea et al. 2009). The only HBRP observed in the nIR is PSR\, J1119$-$6127, with the aim of  finding evidence of  fallback       disc which might have exerted an extra torque on the pulsar, thus modifying its spin-down parameters and increasing the inferred magnetic field value. However,  \vlt\ observations failed in detecting nIR emission from the pulsar position (Mignani et al. 2007c) and did not rule out the presence of a disc with accretion rate $\dot{M} < 3\times 10^{16}$ g s$^{-1}$, still capable of producing a substantial torque on the pulsar.

\section{ Summary and Conclusions}

UVOIR observations not only add a tile to complete the picture of the multi-wavelength phenomenology of different classes of INSs  but they also  play a major role in studying the intrinsic properties of neutron stars and in shedding light on their formation and evolution.  

Like it is shown in the case of RPPs and XDINSs, optical/nUV observations are important, together with the X-ray ones, to build the thermal map of the neutron star surface and to investigate the conductivity in the neutron star interior. nUV observations  are especially important  to study the surface thermal radiation from neutron stars much older than $10^{6}$ years and to determine the temperature of the bulk of the neutron star surface, which is too low  to  be measured in  X-rays where only  very small, hot polar caps are seen (e.g., Becker  2009). This is crucial to constrain the tails of  neutron star cooling models an to investigate possible re-heating processes   (e.g., Page 2009 and  Tsuruta 2009).   While  timing is instrumental in securing the INS identification, the comparison of the UVOIR light curves with the radio, X-ray, and $\gamma$-ray ones is important to  locate different emission regions in the neutron star magnetosphere.  Moreover,  the study of the  giant pulses in RPPs,  so far observed only in the radio and in the optical bands (Shearer et al.  2003), allows to uniquely study the relation between coherent and incoherent emission.  Optical polarisation measurements represent important tests for neutron star magnetosphere models, they allow to measure the neutron star magnetic field and spin axis angles, and to unveil magneto-dynamical interactions between the neutron star and the surrounding PWNe  (see Mignani et al. 2007e). IR observations proved crucial to search for fallback discs  around neutron stars, to verify accretion scenarios, to test magnetars and CCO models, and to investigate the neutron star evolution in the immediate post-supernova phases. Finally,  UVOIR astrometry is still the best way to measure proper motion and parallaxes of radio-silent INSs. Moreover, high-resolution UVOIR imaging  allows to study the morphology and evolution of  PWNe, to search for bow-shocks produced by the INS motion through the ISM  and, thus, to reconstruct its spatial velocity and trace back its birth place and its parental stellar population. 

Table  1  summarizes  the  current UVOIR identification status  for all  classes of  INSs,  i.e. RPPs, XDINSs, and magnetars.   The table also summarises the complete observational data base for all the listed INSs, including multi-band photometry, low-resolution spectroscopy, polarimetry and timing.  As it is seen, a total of 25 INSs have been identified so far, which represents a factor of 3 increase with respect to the number of identifications obtained by the end of the 1990s. However, this is still far from the number of INSs identified in the X-rays (e.g., Becker  2009)  and from the number of  INSs which are being identified in $\gamma$-rays by {\em Fermi} (Abdo et al. 2009, in preparation).  Moreover,  spectroscopy, polarimetry, and timing observations still represent a severe challenge for the  current generation of 8m-class telescopes. This challenge  can be faced with the new generation of  large telescopes, like the 42m european {\em Extremely Large Telescope} ({\em E-ELT}).  Observations with the {\em E-ELT} will open a new era in the UVOIR astronomy of INSs, yielding to at least a hundred of new identification and allowing to carry out studies so far limited to a handful of objects only.

\begin{theacknowledgments}
RPM thanks A. Ibrahim and J. Grindlay for the invitation and the Italian Embassy in Cairo for sponsoring his participation, and acknowledges STFC for support through a Rolling Grant. 
\end{theacknowledgments}


\end{document}